Self-organized criticality of a simplified integrate-and-fire neural model on random and small-world network

Hyung Wooc Choi, Nam Jung, and Jae Woo Lee

Department of Physics, Inha University, Incheon 402-751 Korea

We consider the criticality for firing structures of a simplified integrate-and-fire neural model on the regular network, small-world network, and random networks. We simplify an integrate-and-fire model suggested by Levina, Herrmann and Geisel (LHG). In our model we set up the synaptic strength as a constant value. We observed the power law behaviors of the probability distribution of the avalanche size and the life time of the avalanche. The critical exponents in the small-world network and the random network were the same as those in the fully connected network. However, in the regular one-dimensional ring, the model does not show the critical behaviors. In the simplified LHG model, the short-cuts are crucial role in the self-organized criticality. The simplified LHG model in three types of networks such as the fully connected network, the small-world network, and random network belong to the same universality class.

I. INDRODUCTION

The self-organized criticalities of the neural network and the dynamics of the brain have been greatly attracted in a decade [1-10]. The dynamics of the brain is very complicated. In the brain, each neuron receives firing signals from many nearest neighbors. Beggs and Plenz reported the critical dynamics of the firing patterns in the grown neural networks by the cultured neurons of a young rat [1,2]. They observed the power law of the distribution of the avalanche size with the critical exponent $-1.5$ [1]. Many similar experimental observations reported the critical behaviors of the avalanches in the layers of the cortex [3-6].

In the subcritical state of the neural networks, the input signals die out and are insensitive to the small signals. In the supercritical state the entire network fires by even small inputs. In the critical state, the neural network can manage the small and large inputs, effectively [7]. Kinouchi and Copelli suggested that a network of excitable elements has its sensitivity and dynamic range maximized at the critical point [8]. At criticality, a network of excitable elements can be extremely sensitive to the small perturbations and still able to detect large input without saturation [7-9].

The critical behaviors of the neural avalanche agree to the paradigm of the self-organized criticality (SOC) [10-13]. Levina *et al* proposed an integrated-and-fire model (called the LHG model) with short-term plasticity. In a fully connected neural network, they observed the SOC and obtained the critical exponents of the avalanche [9]. The critical exponents of the LHG were equal to the exponents of the mean-field model. Li and Small observed the neural avalanches of a self-organized neural network with active-neuron-dominant structure [14]. They reported that the process of network learning by spike-timing dependent plasticity increases the complexity of network structure. Peng and Beggs proposed a cellular automaton model of neural networks that combines short-term synaptic plasticity and long-term plasticity [15]. In this work we consider a simplified integrate-and-fore model. We simplified the LHG model. In this model we consider the constant synaptic strength among the neural connections. We observed the dynamics of the avalanche on the regular network, the small-world network, and the random networks. In section 2 we introduced the simplified integrate-and-fire neural model. In section 3 we presented the numerical results. We gave the concluding remarks in section 4.

## II. SIMPLIFIED INTEGRATE-AND-FIRE-NEURAL MODEL

The brain is composed by huge numbers of neurons. They are connected by very complex networks. The firing patterns of the neural network are described by the many neural models. Here, we adapt a simplified integrate-and-fire neural model. Levina *et al* introduced an integrate-and-fire-neural model (LHG model) for a fully connected neural network with N neurons such as [9]

$$\frac{\partial V_i}{\partial t} = \delta(t - t_{dr}^i)I^{ext} + \frac{1}{N-1}\sum_{j=1}^{N-1} uJ_{ij}\delta(t - t_{sp}^j) - V_{max}\delta(t - t_{sp}^i), \qquad (1)$$

$$\frac{\partial J_{ij}}{\partial t} = \frac{1}{\tau_J}\left(\frac{\alpha}{u} - J_{ij}\right) - uJ_{ij}\delta(t - t_{sp}^i), \qquad (2)$$

where $V_i$ is a membrane potential of a neuron and $J_{ij}$ is the strength of the synaptic connection between neuron i and neuron j which characterized by the neurotransmitters. Each neuron is randomly driving by the external noise with the strength $I^{ext}$ at a random driving time $t_{dr}^i$ for the neuron i. $V_{max}$ is a threshold potential, and $t_{sp}^i$ is a spiking time of the neuron i when the membrane potential exceeds the threshold potential. $\tau_J$ is the relaxation time of the synaptic strength, $\alpha$ is a control parameter that effects some target values of synaptic recovery, and u is saturation constant of the synaptic strength. The LHG model was simulated on the fully connected neural network and observed the power-law distribution of the avalanche [9]. We consider a simplified LHG (SLHG) model to observe the main effects of the self-organized criticality in the firing dynamics of the neural networks [16]. In our model we set the synaptic strength as a constant, J = constant. Therefore, we don't need equation (2). We choose the control parameter as u = 0.5, $V_{max}$ = 1, and N=300. We solve the equation (1) by the fourth order Lunge-Kutta method. We consider the one-dimensional ring network, random network and small-world network as a neural connecting network.

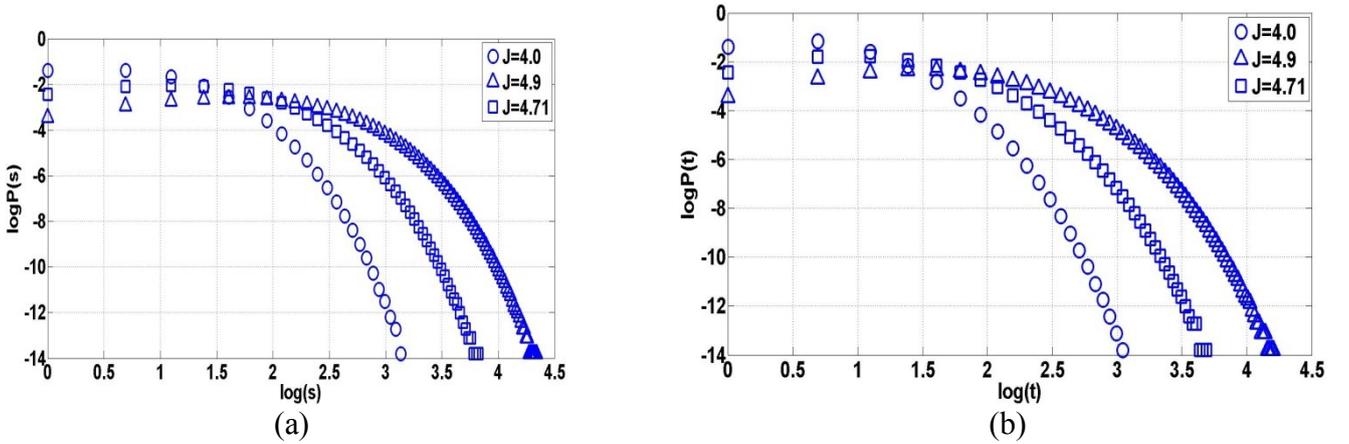

(a)            (b)

Fig. 1. Probability distribution function of (a) the avalanche size and (b) the life time for the avalanche in one dimensional lattice with the periodic boundary condition.

III. SIMULATION RESULTS

We generate one-dimensional neural networks with periodic boundary condition, small-world networks, and random networks. We investigate short-cut effects of the simplified LHG model on the avalanche structures. At first, we consider a one-dimensional ring connection of the neurons. In the previous studies, we observed the critical synaptic strength as $J_c = 4.71$ in the fully connected graph [16]. By the random input, neurons are approaching to the tipping points. One of them reaches the firing membrane potential. This firing triggers the subsequent firing of the neighboring neurons. The whole numbers of the fired neurons are belonging to an avalanching cluster. We measure the distribution of the avalanche size and the life time of the avalanche when the system reaches to the state-state. The life time of an avalanche is defined by the elapsed time from the ignited firing to the terminating of the avalanche. In Fig. 1 we represent the size distribution of the avalanche and the distribution of the avalanche life time. We observed that there are no critical behaviors of the avalanche in the one-dimensional lattice. The distribution functions decay exponentially regardless of the synaptic strength. There was no power laws when we adjust the synaptic strength up to $J = 5$.

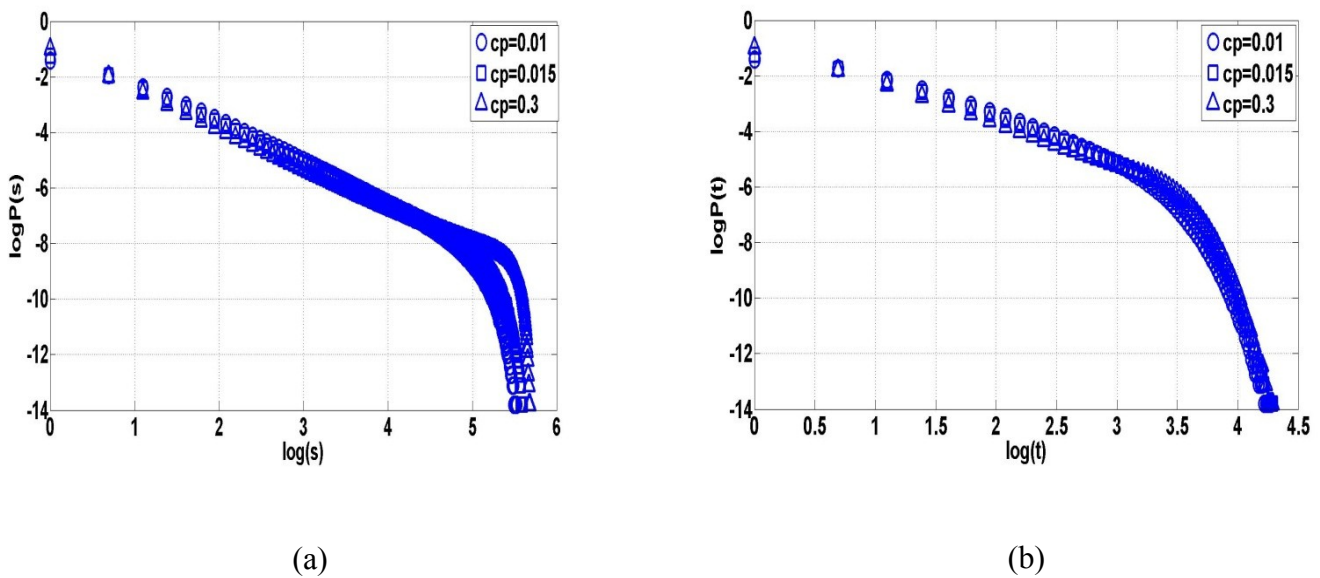

(a)          (b)

Fig. 2 Probability distribution function of (a) the avalanche size and (b) the life time for the avalanche in the small-world networks. We increase the short-cut probabilities from cp=0.01 to cp=0.3. We observed the power-law distribution of the distribution.

To observe the short-cut effects in SLHG model, we consider the small-world networks. We generated the Newman-Watts type small world networks [17-20]. When we generate the small world network, we start a regular one-dimensional ring with the coordination number 4. We inset a short-cut between two nodes with the short-cut probability p. As the fully connected network, we observed the power-law behavior for the distribution of the avalanche and the life time. We observed the critical synaptic strength as $J_c = 4.71$ regardless of the short-cut probability. We plot the distribution of the avalanche size in Fig. 2 (a) and the distribution of the avalanche life time in Fig. 2 (b) at the synaptic strength $J = 4.71$. We observed the power-law behavior for both the distribution from the short-cut probability p=0.01 to p=0.3. For the small short-cut probability, the system shows the self-organized criticality. The short-cut is crucial role of the power-law in the SLHG model. The distribution of the avalanche size follows the power-law like $P(s) \sim s^{-\alpha}$. We obtained the critical exponent as $\alpha = 1.41(5)$ by the least-square fit. The critical exponent $\alpha$ is equal to that of the fully connected network [16]. The distribution of the avalanche life time also shows the power law, $P(t) \sim t^{-\beta}$. We obtained the exponent of life time distribution as $\beta = 1.50(9)$. These critical exponents are same as those obtained from the fully-connected network [16].

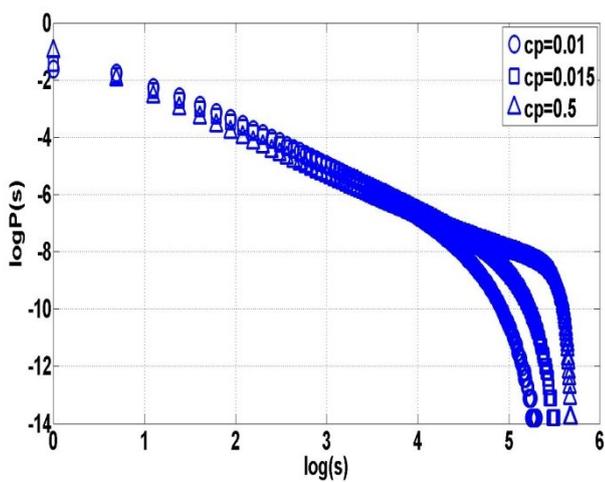
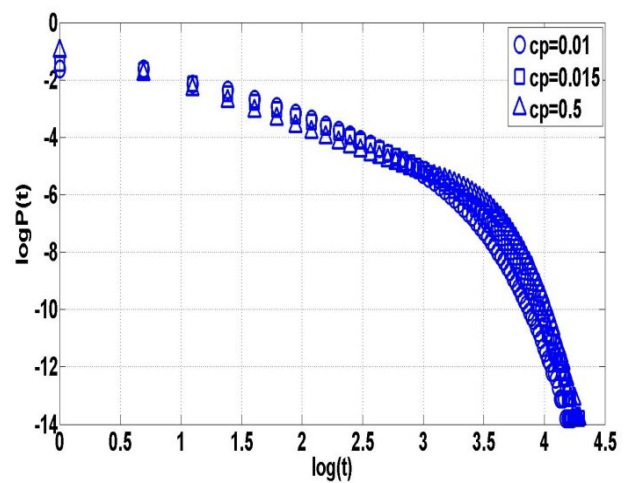

(a)

(b)

Fig. 3. Probability distribution function of (a) the avalanche size and (b) the life time for the avalanche in the random networks. We generate the random network with the random connecting probabilities from cp=0.01 to cp=0.5. We observed the power-law distribution of the distribution.

In Fig. 3 we present the probability distribution of the avalanche size and the life time in the random networks [18-20]. We generate random networks changing the connection probability from cp=0.01 to cp=0.5. We observed the similar probability distribution function like the case of the small-world networks. The distribution of the avalanche size and the life time shows the power law with the same exponent as the small-world network and fully-connected networks. Only cut-off size and cut-off time depend on the connection probability of the random network. Therefore, we conclude that the criticality of the SLHG model does not depend on the substrate networks. However, the long-range connections such as the short-cut are crucial role in the criticality of this model. Three types of networks such as fully-connected network, small-world networks, and random network are belonging to the same universality class in the SLHG model.

IV. CONCLUSIONS

We consider the SLHG neural model on the three different kinds of the networks. We observed the power law behaviors of the probability distributions for the avalanche size and the life time of the avalanche in the small-world networks and the random networks. However, in the regular network, the SLHG model does not show the criticality. The shut-cuts in the networks are crucial role in the self-organized criticality of the SLHG model. We conclude that the SLHG model on the three kinds of network such as the fully-connected network, the small-world network, and the random network belong to the same universality class.

ACKNOWLEDGEMENTS

This research has been supported by the research fund of Inha University.